\documentclass[conference]{IEEEtran}

\usepackage{amsmath,amssymb}
\usepackage{algpseudocode}
\usepackage{array}
\usepackage{booktabs}
\usepackage{cite}
\usepackage{float}
\usepackage{graphicx}
\usepackage{microtype}
\usepackage{tikz}
\usepackage{xcolor}
\usetikzlibrary{arrows.meta,positioning,fit,calc}
\usepackage[hidelinks]{hyperref}
\usepackage{orcidlink}

\definecolor{methodblue}{RGB}{0,114,178}
\definecolor{methodorange}{RGB}{213,94,0}
\definecolor{lightblue}{RGB}{222,235,247}
\definecolor{lightgray}{RGB}{238,238,238}

\title{Certifying Repeatable Entry-Replay Witnesses\\
for COBOL-to-Java Migration}

\author{
\IEEEauthorblockN{Andras L. Ferenczi\,\orcidlink{0000-0001-6785-9416},
Mariya Bessonov, Allwyn Ponnudurai\\
Chandrakanth R. Mamillapalli, Krishna K. Maddula, Krishna K. Lingamneni}
\IEEEauthorblockA{American Express}
}

\begin{document}
\maketitle

\begin{abstract}
Producing synthetic data for comparison testing a migrated COBOL application is a major challenge that agentic migrations face today. Our prior paper, the \emph{Agentic Method for Deterministic Validation of Legacy Code Migration}, presents a deterministic oracle to facilitate such migrations, the \emph{Locksmith Loop}. We now propose a novel method for achieving high branch and paragraph coverage through generated synthetic end-to-end test cases, by means of \emph{entry-replay test-case certification} by observing the code execution trace on a surrogate COBOL program that mocks external interactions via a chronological external-operation tape.
We applied our method to a very large COBOL program with 33,429 lines of code to generate 849 synthetic end-to-end test cases covering 2,973/3,806 source branches (78.1\%) and 989/1,033 main-AST paragraphs (95.7\%).
Using the Locksmith Loop's projected-agreement validation, the generated Java code matched the COBOL surrogate on the specified observations across all 849 certified synthetic end-to-end test cases.
During the campaign we detected 61 defects and fixed the Java generator accordingly. To test whether the coding agent could silently introduce business logic bugs, we seeded 450 defects into the Java code, such as deleted writes, deleted paragraph calls, and altered rounding, to verify how many of these would be detected by the gating methodology. The gate detected 47.6\%, including all deleted writes and paragraph calls in executed code, but was unable to detect the rounding and precision defects.
\end{abstract}


\section{Introduction}

Agentic legacy migration projects need deterministic methods for validating migrated behavior. Frequently, lack of documentation and test data makes it difficult to validate parity between the legacy code (COBOL, in our case) and the new migrated Java code. Unit tests go only so far as to validate local functionality, and integration tests can go only so far as to minimally validate functionality. Even with comprehensive quality assurance user testing, there is a limited guarantee that the system will work as expected with the COBOL code being the ground truth.

The Locksmith Loop~\cite{ferenczi2026locksmith} provides two methods for generating witnesses (synthetic end-to-end test cases): the Witness Search (input-space exploration) and the Mutation (parity-preserving code mutation).
The Locksmith Loop provided a witness generation method demonstrated on a 4,114-line, 142-paragraph COBOL codebase. This paper evaluates a methodological successor to the Locksmith Loop on a 33,429-line, 1,033-paragraph program.

We focus in this research on a novel approach for witness discovery while reusing the Locksmith Loop agentic method for checking projected agreement between the migrated Java code and the surrogate COBOL codebase on certified witnesses. It is important to clarify that the method does not guarantee production functional parity. It is nevertheless helpful for discovering behavioral differences that might otherwise not be identified.

We ask:
\textbf{RQ1:} can our method grow the coverage of the COBOL surrogate, and how does coverage accumulate as the certified catalog grows?
\textbf{RQ2:} how does the source-outcome set of the final certified catalog differ from that of the retained local-search corpus?
\textbf{RQ3:} how many seeded translation defects does the certified gate detect under the observation projection $\Phi_H$ (Section~\ref{sec:method}), and where does detection fail?
\textbf{RQ4:} what kinds of COBOL-to-Java migration bugs did the repeatable certification method expose, and how many repairs did each kind require?
\textbf{RQ5:} what are the guarantees and the limits of our witness certification process?

\section{Related Work}

At a high level, we reuse the Locksmith Loop~\cite{ferenczi2026locksmith} approach for validation against a deterministic oracle, including mocked interactions with external systems such as databases, message queues, and other programs, but not its method for proposing witnesses. The local-state adaptation is related to testability transformation~\cite{harman2004testability} and to DART, Driller, and T-Fuzz~\cite{godefroid2005dart,stephens2016driller,peng2018tfuzz} (see~\cite{baldoni2018symbolic} for a survey). We use agentic functionality for this task and count only end-to-end source replay as coverage. Elbaum et al.~\cite{elbaum2006carving} and Saff et al.~\cite{saff2005factoring} extract intermediate program states from an execution and use those to replay a subcomponent of the program in isolation; we do exactly the opposite, proposing witnesses in isolation and walking our way upstream to the conditions that satisfy execution of the subcomponent.
Kumar et al.~\cite{kumar2024validation} and Hans et al.~\cite{hans2025testing} use symbolic execution on the COBOL source code to generate JUnit test cases, which are then run against the Java code. Similarly to the Locksmith Loop, this method implies like-for-like migration rather than object-oriented refactoring. However, their method is performed strictly at the unit test level, with the known limitation of path explosion, thus making it intractable for very large programs. Etemadi et al.~\cite{etemadi2025mokav} begin with functioning test inputs that produce matched behavior in two programs, then use an LLM iteratively to generate new inputs until it identifies different behavior between these (disagreement as an oracle). A disagreement shows that the two programs differ, but not which one is correct. As an alternative for settings lacking a deterministic ground truth, Chen et al.~\cite{chen2018metamorphic} use metamorphic testing, a method that verifies relations that must exist between multiple execution outputs. While this can indeed identify discrepancies between original and migrated code, it only detects violations of the chosen relations, and so does not offer a comprehensive solution. Pnueli et al.~\cite{pnueli1998translation} prove translation accuracy by establishing a formal correspondence between the source and target programs and verifying it for each translation run. In contrast, our method uses an empirical approach, which we demonstrate to be feasible at the scale of a 33,429-line program using agentic coding methods. Our method addresses the frequently most realistic scenarios where suitable from-entry source tests are unavailable.
Dijkstra's 1975 paper~\cite{dijkstra1975guarded} presents a formal method for defining the weakest execution precondition for a guarded command in order for the execution to result in expected behavior. By decomposing the program into such guarded commands one can better derive the end-to-end execution preconditions. In contrast, our method works on an already existing COBOL program, computes the requirements empirically and produces a precondition sufficient for a specific witness to work. Chv{\'a}tal~\cite{chvatal1979setcover} is a foundational method for a greedy subset (witness) search by repeatedly picking the subset with the lowest cost per newly covered element in every iteration and is repeated until all elements are found, a technique also applied to test-suite minimization~\cite{harrold1993test} and directly applicable to reducing our own certified witness catalog.

\section{Source-Replay-Certified Witness Method}\label{sec:method}

\begin{figure}[b]
\centering
\begin{tikzpicture}[
  font=\footnotesize,
  node distance=5.5mm,
  stage/.style={draw, rounded corners=1.2pt, align=left,
    text width=.86\columnwidth, minimum height=16mm,
    inner xsep=2.6mm, inner ysep=2.6mm},
  build/.style={stage, fill=lightgray},
  gate/.style={stage, fill=lightblue, draw=methodblue},
  line/.style={-{Latex[length=1.5mm]}, line width=.45pt}
]
\node[build] (construct) {
  \textbf{1. Construct candidate (no coverage).}\\
  Observe concrete transitions. Lift them toward entry.\\
  \emph{Output:} witness $w=(\sigma_0,\tau)$ or a named refusal.};
\node[gate, below=of construct] (source) {
  \textbf{2. Certify on compiled source.}\\
  Reset the complete harness and replay $w$.\\
  \emph{Pass:} stable source signature and conforming tape.\\
  \emph{Output:} hash-bound certificate and coverage credit.};
\node[gate, below=of source] (target) {
  \textbf{3. Validate generated Java.}\\
  Replay the unchanged certified witness.\\
  \emph{Pass:} Java matches the declared source projection.\\
  Target-only repairs return to this gate.};
\draw[line] (construct) -- node[right, font=\scriptsize] {candidate} (source);
\draw[line] (source) -- node[right, font=\scriptsize] {certificate} (target);
\end{tikzpicture}
\caption{The workflow has three noninterchangeable claims: candidate
construction, source certification, and target agreement.}\label{fig:method}
\end{figure}
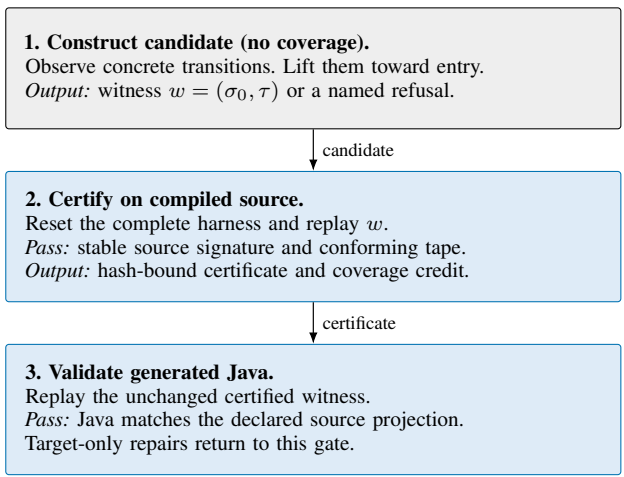

We define $H$ as the frozen harness (rulebook) that guides the end-to-end process. This includes the rules for resetting the working storage and database to their initial state (for example, the initial 500 records) so that witnesses all see the same exact setup, the rules for which entry values a witness may set and where, how external operation calls are matched against the tape and overlaid onto physical record layouts, what gets instrumented and counted as in scope, what counts as an accepted terminal outcome, and the normalizers used when comparing runs. Figure~\ref{fig:method} shows the steps involved in the process, including: candidate construction, source certification, and target validation.
We also define a witness as $w=(\sigma_0,\tau)$, where $\sigma_0$ is a list
of permitted program entry values and $\tau$ is the chronological tape of
operation keys paired with supplied outcomes from external interactions
(database, including IMS DL/I segment calls, message queue, file transfer, etc.).
We evaluate a run as $E=(B,r,a,D,Q,O,F,S,J)$, with $B$ being the set of
branch directions that include \texttt{IF}, \texttt{EVALUATE},
\texttt{PERFORM UNTIL}, \texttt{SEARCH}, and conditional handlers; $r$ is
the return code of the program (for example, \texttt{RETURN-CODE = 0});
$a$ is whether the program abended (e.g., \texttt{INVALID KEY} /
\texttt{SIZE ERROR}); $D$ is the display (console) output; $Q$ is the
paragraph trace (recorded upon entry into a paragraph); $O$ is the trace
of the external calls, such as a database call, message queue
interaction, a subprogram \texttt{CALL}, etc. This is helpful for
verification of behavior parity mismatches, such as translation bugs
that can be missed in the final state. $F$ is the final dataset state;
the state of the dataset following the witness propagation has to match
between the compared codebases. $S$ is the declared final program state (the value of the variables at the moment the program terminates); $J$ is the list of consumed tape indices.
\begin{equation}
\begin{aligned}
\Phi_H(E)={}&(r,a,N_D(D),N_Q(\operatorname{first}(Q)),\\
             &\quad N_O(O),F,S,J),\\
\operatorname{Agree}_H(w)\iff{}&
\Phi_H(E_S)=\Phi_H(E_J).
\end{aligned}
\label{eq:agreement}
\end{equation}
In Eq.~\eqref{eq:agreement}, $E_S$ and $E_J$ are the certified COBOL run and the Java run of the same witness $w$, and $\Phi_H(E)$ is the normalized comparison
projection, accounting for syntax differences between the two
programming languages. It drops $B$, since COBOL's branching structure
has no faithful Java counterpart; keeps $r$ and $a$ as-is (constant here, since only zero-return executions are certified); and
normalizes $D$ (for cosmetic formatting differences),
$Q$ reduced to first-occurrence order (for structural translation differences),
and $O$ (for naming-convention differences between COBOL and Java call
identifiers). $F$, $S$, and $J$ are compared exactly, without normalization. Although $B$ is not compared, it is used for source certification and coverage accounting in
Eq.~\eqref{eq:coverage}.
Let $U_H$ be the emitted source-direction universe and $I_H$ its reviewed
exclusions. Only fresh certified source runs contribute $B_C^H(w)$, the
certified branch set for witness $w$:
\begin{equation}
\operatorname{cov}_H(W)=
\frac{\left|\left(\bigcup_{w\in W}B_C^H(w)\right)
\cap(U_H\setminus I_H)\right|}
{|U_H\setminus I_H|}.
\label{eq:coverage}
\end{equation}
Thus full scoped coverage, source certification, and target agreement are
three distinct claims.

\looseness=1
The procedure in Fig.~\ref{fig:construct-certify} takes as inputs a goal $g$, such as a source direction/branch outcome and $p$, the paragraph in which the goal lives. $T$ is a table with each row representing entries for a recovered paragraph unit, $Q_q$, which is a COBOL paragraph plus all its internal calls to other
paragraphs via \texttt{PERFORM}/\texttt{PERFORM THRU}. A row in $T$
will have the input state (e.g., \texttt{ITEM-ID}, \texttt{WAREHOUSE-CODE}),
the local tape (e.g., $[1]$ \texttt{DB2-LOOKUP:7734} $\to$
$\{\texttt{STOCK-QTY}: 120, \texttt{STATUS}: \text{``IN-STOCK''}\}$),
observed directions (e.g., \texttt{2100-CHECK-INVENTORY}:
\texttt{IF WS-QTY < WS-ORDER-QTY} $\to$ \texttt{TRUE}), executed edges
(e.g., \texttt{2000-VALIDATE-ORDER} $\to$ \texttt{2100-CHECK-INVENTORY}
(\texttt{PERFORM}); \texttt{2100-CHECK-INVENTORY} $\to$
\texttt{2110-LOOKUP-ITEM} (\texttt{PERFORM})), successor entry state
(e.g., \texttt{WS-VALIDATION-RESULT = "REJECTED-OUT-OF-STOCK"}), and a
backpointer (e.g., $\to$ Row \#4182, the row this exploration run was
chained from). Initially, $T$ starts empty and the tool makes educated guesses by looking at the code (e.g., IF-statements) plus data that existing witnesses have. The guesses that actually work are inserted into $T$.
We then check at line 3 whether we get rows in $T$ that fire the goal. For example, $p = \texttt{2100-CHECK-INVENTORY}$ and $g$ is ``the \texttt{IF WS-QTY < WS-ORDER-QTY} check evaluates \texttt{TRUE}'' (i.e., the out-of-stock branch). Lines 8 to 16 are doing the backward-lifting walk through paragraphs until it reaches the entry point. At line 10, \texttt{DEMANDOBSERVE} makes an attempt to fetch the rows satisfying the condition, however, if none found, it will attempt to generate new rows to meet the ask. Line 14 picks one row from the candidates satisfying edge $e$, conditioned by an $H$ policy, and prepends it to the $\beta$ list, and in line 15 $R$ gets updated with whatever requirements the upstream step needs. Also, now $p$ is updated to $q$ for the next iteration.
Lines 17 to 25 are the refinement loop. It runs up to $H.\mathit{refineCap}$, a fixed cap. Based on $\beta$, \texttt{MATERIALIZE} produces on line 18 a candidate witness
consisting of the entry input values (drawn from the earliest link in
the chain, informed by $R$) and a single sequential tape stitched
together from each row's local tape/environment data along the chain. Next, on line 19 \texttt{REPLAY} runs the compiled COBOL surrogate ($P_C$) injecting $w$'s entry values while consuming $w$'s tape, producing the COBOL execution $E_c$, which is then retained at line 20. Lines 21 to 23 break from the loop if $E_c$ is accepted per the harness rules and branches covered by $E_c$ include the original goal, $g$. \texttt{REFINE} at line 24 analyzes the failure of the current execution (using
the fresh evidence in $E_c$, together with the accumulated table $T$)
and re-defines the requirement $R$ and the chain $\beta$ for the next
attempt. Line 26 tests whether the goal is absent from the branch set of the final refinement attempt, and if so line 27 returns the refusal, regardless of whether earlier attempts in the loop might have hit the goal but failed the acceptance check on that particular iteration. The \texttt{FRESHREPLAY} at line 29 is necessary to ensure that the witness still meets the goal under a fresh replay ($E_f$) in which $H$ is reset to initial condition. Line 30 verifies that the witness reproduces the same behavior on this
second, independent run as it did during refinement --- not merely
that both runs terminated successfully. \texttt{SOURCESIG}
captures all deterministic elements of $E$ except $J$ (which is
checked separately via \texttt{TAPEOK} on line 31):
$(B, r, a, D, Q, O, F, S)$. Line 33 returns a certification (hash-bound to $w$ and $E_f$) covering
the compiled surrogate, harness configuration, and the witness itself.
Should any of these change afterward, the prior certification is
nullified; a mismatch signals that the other witnesses also have
to be re-verified. Procedure \texttt{NEXTOUTCOME} in
Fig.~\ref{fig:tape-matcher} scans row $M$'s tape forward from the
last consumed index to find the next entry applicable to external
call $c$, under harness $H$; it marks that entry used and returns its
overlaid outcome, or fails if no applicable entry remains.
\begin{figure}[t]
\footnotesize
\algrenewcommand\algorithmicindent{1em}
\begin{algorithmic}[1]
\Procedure{ConstructAndCertify}{$g,p$}
  \State $T\gets\Call{SeedAndForwardClose}{H}$
  \State $R\gets\Call{RowsFiring}{T[p],g}$
  \If{$R=\emptyset$}
    \State \Return \Call{Refuse}{\textsc{no-firing}}
  \EndIf
  \State $\beta\gets[\,]$
  \While{$p\ne\textsc{entry}$}
    \State $e=(q\!\to\!p)\gets\Call{ChoosePredecessor}{p,H}$
    \State $M\gets\Call{DemandObserve}{q,e,R,T,H}$
    \If{$M=\emptyset$}
      \State \Return \Call{ClassifyRefusal}{q,e,R}
    \EndIf
    \State $\beta\gets[(e,\Call{ChooseRow}{M,H})]\mathbin{\|}\beta$
    \State $R\gets\Call{RequiredInputs}{M,q}$; $p\gets q$
  \EndWhile
  \For{$n\gets1$ to $H.\mathit{refineCap}$}
    \State $w\gets\Call{Materialize}{R,\beta,H}$
    \State $E_c\gets\Call{Replay}{P_C,w,H}$
    \State \Call{RetainObservations}{$T,E_c$}
    \If{\Call{Accepted}{$E_c,H$} and $g\in E_c.B$}
      \State \textbf{break}
    \EndIf
    \State $(R,\beta)\gets\Call{Refine}{R,\beta,E_c,T,H}$
  \EndFor
  \If{$g\notin E_c.B$}
    \State \Return \Call{Refuse}{\textsc{refinement-cap}}
  \EndIf
  \State $E_f\gets\Call{FreshReplay}{P_C,w,H}$
  \State $stable\gets
    (\Call{SourceSig}{E_c,H}=\Call{SourceSig}{E_f,H})$
  \State $tapeOK\gets\Call{TapeOK}{E_c,w,H}\land
    \Call{TapeOK}{E_f,w,H}$
  \If{\Call{Accepted}{$E_f,H$} and $stable$ and $tapeOK$}
    \State \Return \Call{HashBind}{$w,E_f$}
      \Comment{credit only $E_f.B$}
  \EndIf
  \State \Return \Call{Reject}{\textsc{source-certificate}}
\EndProcedure
\end{algorithmic}
\caption{Witness construction and source certification. Internal runs add
facts but never coverage; only the fresh replay earns credit.}
\label{fig:construct-certify}
\end{figure}

\looseness=1
Per Fig.~\ref{fig:construct-certify}, the flow entails finding a row in $T$ that fires the branch when going after it. What is not explicitly shown here is that the agent may also take a certified witness and tweak its entry values or the tape to go after an uncovered branch. Skipping the $T$ lookup does not mean skipping any of the checks: such candidates still face the same replay, fresh-replay, stability, and TAPEOK gates before earning coverage credit.

\begin{figure}[t]
\footnotesize
\algrenewcommand\algorithmicindent{1em}
\begin{algorithmic}[1]
\Procedure{NextOutcome}{$c,M,H$}
  \For{$i\gets M.last+1$ to $|M.tape|$}
    \If{\Call{Applicable}{$H,c,M.tape[i].key$}}
      \State $M.last\gets i$; append $i$ to $M.used$
      \State \Return \Call{Overlay}{$c,M.tape[i].outcome$}
    \EndIf
  \EndFor
  \State \Return \Call{Fail}{\textsc{missing-match}}
\EndProcedure
\end{algorithmic}
\caption{Chronological external-outcome tape matching. Matching moves forward;
a missing applicable entry fails replay.}\label{fig:tape-matcher}
\end{figure}

\section{Evaluation}\label{sec:eval}

\subsection{Setup}\label{sec:setup}
As in the Locksmith Loop~\cite{ferenczi2026locksmith}, we used a production-like COBOL program, here eight times larger (33,429 lines). We instrumented the source's decision constructs and mocked all external interactions (mediated files, database operations, subprograms), correctly handling binary, packed-decimal, redefined, and repeated record layouts along the way. We compiled it using GnuCOBOL 3.2.0 on an Arm workstation, targeting OpenJDK 17. Unlike the Locksmith Loop, we did not leverage any test data, but instead generated inputs based on the COPYBOOK structures used by the program.\footnote{No customer data were used in any experiment during the campaign.}

$F$ holds the final contents of the output files generated by each run in the certified runtime. The remaining output files are served by the mocked I/O layer, so $J$ and $O$ check their operations but not their payloads.

\emph{Agent governance.} We used commercial LLM-based coding agents to propose witnesses and repairs. The models and agent harnesses were upgraded several times as new versions became available during the campaign. Certification, replay, tape matching, and the $\Phi_H$ comparison are deterministic code that every agent output must pass. An author reviewed the repairs in the surrogate, mock, and capture oracle and accepted them only after the full catalog re-certified.

Sections~\ref{sec:coverage}--\ref{sec:ledger} answer RQ1--RQ4, Section~\ref{sec:cost} reports cost, and Section~\ref{sec:discussion} answers RQ5.

\subsection{Coverage (RQ1)}\label{sec:coverage}
In the end we covered 2,973/3,806 source branches (78.1\%), and 989/1,033
main-AST paragraphs (95.7\%). This does not mean that either of the
uncovered paragraphs or branches is unreachable, but rather reflects
the fact that we intentionally stopped construction once main-AST paragraph coverage passed a 95\% target. Even with a ``perfect'' deterministic match under the declared observation projection, there could exist undetected subtle defects hiding in the code, as quantified in Section~\ref{sec:faults} and Table~\ref{tab:projection}. For example, the Java code might write the wrong value somewhere in the middle of the program run but then overwrite it with the correct value. Table~\ref{tab:measurements} reports the full
case measurements.

\begin{figure}[t]
\centering
\includegraphics[width=\columnwidth]{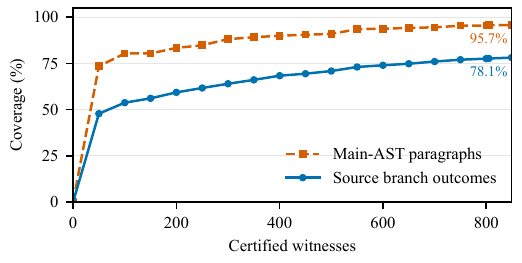}
\caption{Certified coverage as the catalog grows, measured every 50 witnesses and at 805 and 849.}\label{fig:growth}
\end{figure}

\begin{figure*}[!t]
\centering
\includegraphics[width=.68\textwidth]{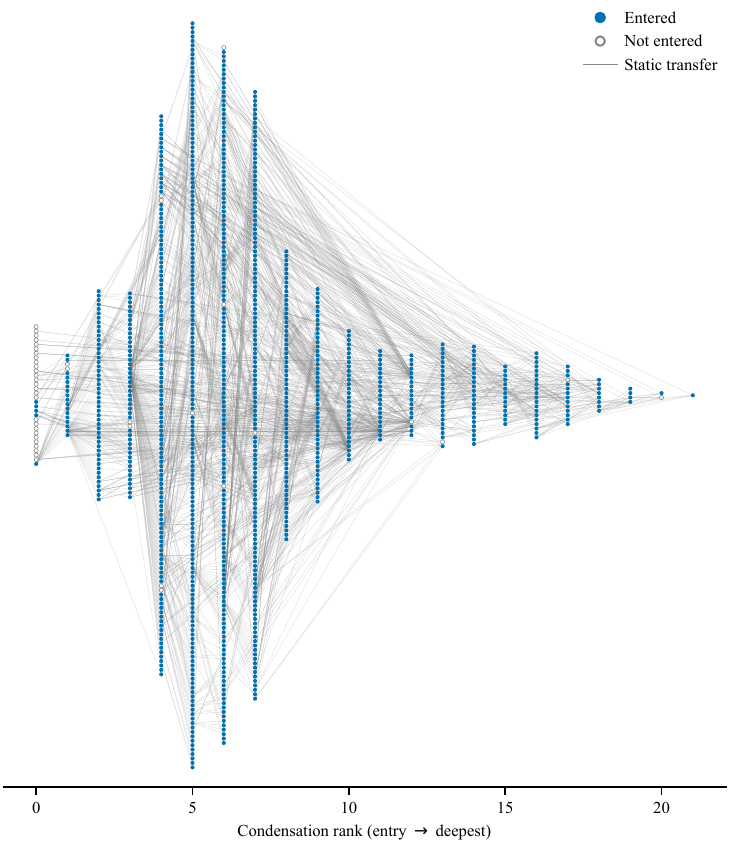}
\caption{Scale overview of the recovered graph: 1,033 main-AST paragraphs and
2,029 resolved transfers. Nodes are arranged by longest-path rank in the
strongly connected component condensation; filled nodes are the 989 paragraphs
entered by certified witnesses. The dense view shows aggregate structure, not
evidence about any individual path.}\label{fig:dag}
\end{figure*}

\begin{figure}[tb]
\centering
\includegraphics[width=\columnwidth]{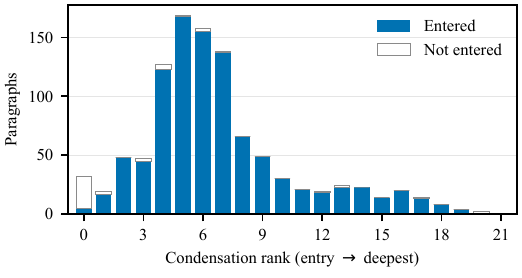}
\caption{Entered and unentered main-AST paragraphs by longest-path rank in
the recovered graph's strongly connected component condensation. Coverage is
not a shallow prefix; unentered paragraphs occur throughout the depth.}\label{fig:depth}
\end{figure}

Certified branch and paragraph coverage reached 47.8\% and 73.5\% after 50 witnesses, which represent only 6\% of the catalog. We attribute this to shared validation paragraphs and common error handling routines. Coverage rose to 68.3\% branch and 89.9\% paragraph after 400 witnesses, and to 78.1\% and 95.7\% after 849 (Fig.~\ref{fig:growth}). Figures~\ref{fig:dag} and~\ref{fig:depth} show how the final paragraph coverage is distributed across the recovered graph and its depth ranks.

Of the 44 unentered main-AST paragraphs, 27 occur at condensation rank zero, making this the largest remaining structural concentration.

The other 17 unentered paragraphs are sparsely distributed through ranks 1--20, rather than forming a shallow coverage cutoff.

\begin{figure}[!htbp]
\centering
\includegraphics[width=\columnwidth]{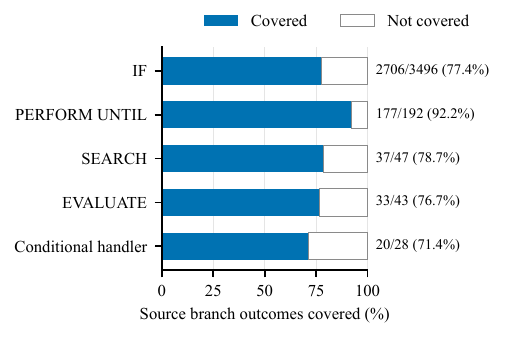}
\caption{Certified branch coverage by source decision construct.
The aggregate is 2,973/3,806 outcomes (78.1\%) across 849 entry-replay-certified
witnesses.}\label{fig:expanded-coverage}
\end{figure}

Figure~\ref{fig:expanded-coverage} decomposes the aggregate certified branch
coverage in Table~\ref{tab:measurements} by source decision construct.

\texttt{PERFORM UNTIL} outcomes had the highest measured coverage (92.2\%), while conditional handlers had the lowest (71.4\%); \texttt{IF}, \texttt{SEARCH}, and \texttt{EVALUATE} remained between 76.7\% and 78.7\%.

\subsection{Corpus Comparison (RQ2)}\label{sec:corpus}
The agent persisted 133,261 distinct isolated observation rows, most of them redundant or from failed attempts. Separately, 872 whole local/boundary cases were retained as the frozen reporting corpus, covering 2,860 branch outcomes. The certified catalog covered 2,973, with 2,505 shared between them; 468 appear only in the certified catalog and 355 only in the frozen corpus.

\begin{table}[t]
\caption{
Measured runtime of the final 849-witness validation gate. This
measurement excludes witness construction, source compilation, Java
generation, and Java compilation and is not an end-to-end
construction-cost measurement.
}\label{tab:cost}
\centering
\footnotesize
\begin{tabular}{@{}p{0.50\columnwidth}rr@{}}
\toprule
Stage & Runs & Wall clock \\
\midrule
Fresh source replay and capture & 849 & 239.826\,s \\
Precompiled Java catalog replay & 849 & 502.846\,s \\
\midrule
Combined measured validation gate & 1,698 & 742.672\,s \\
\bottomrule
\end{tabular}
\end{table}

\begin{table}[t]
\caption{Observed projection signatures and scope across 849 witnesses.
Branch profiles are reporting-only.}\label{tab:projection}
\centering
\scriptsize
\begin{tabular}{@{}>{\raggedright\arraybackslash}p{0.76\columnwidth}>{\raggedleft\arraybackslash}p{0.18\columnwidth}@{}}
\toprule
Measure & Value \\
\midrule
Distinct source branch profiles (not compared) & 807 \\
Distinct combined compared projections & 514 \\
Distinct final output-dataset images & 316 \\
Distinct external-operation key sequences & 306 \\
Distinct first-occurrence paragraph traces & 322 \\
Distinct normalized display sequences & 179 \\
Distinct return/abend outcomes & 1 \\
Unpaired write/rewrite payloads & 36,362 \\
Calls with uncompared request arguments & 44,993 \\
\bottomrule
\end{tabular}
\end{table}

\begin{table}[t]
\caption{Frozen case-study measurements. Agent rows are descriptive workload
accounting only.}\label{tab:measurements}
\centering
\footnotesize
\begin{tabular}{@{}>{\raggedright\arraybackslash}p{0.66\columnwidth}>{\raggedleft\arraybackslash}p{0.28\columnwidth}@{}}
\toprule
Measure & Value \\
\midrule
COBOL source lines & 33,429 \\
Unique main-AST paragraphs & 1,033 \\
Emitted source branch outcomes & 3,806 \\
Covered source branch outcomes & 2,973/3,806 \\
Certified witnesses & 849 \\
Source-replay checks passed & 849/849 \\
Entered main-AST paragraphs & 989/1,033 \\
Entered procedure-copybook labels & 88/131 \\
Unentered with a resolved static caller & 17 (7 non-\texttt{EXIT}) \\
Unentered with no resolved static caller & 27 (20 non-\texttt{EXIT}) \\
Agent requests & 1,067 \\
Agent new tokens & 203\,M \\
Certified operation-tape entries & 199,906 \\
External operations in the snapshots & 199,906 \\
Certified tapes matching observed trace length & 849/849 \\
Observed operations/witness (median; IQR; max) & 244; 209--280; 477 \\
Interleaved traces / immediate $A$--$B$--$A$ & 849/849; 799/849 \\
Certified source branch coverage & 2,973/3,806 (78.1\%) \\
Certified main-AST paragraph coverage & 989/1,033 (95.7\%) \\
Java agreement under $\Phi_H$ & 849/849 \\
\bottomrule
\end{tabular}
\end{table}

\begin{figure}[t]
\centering
\includegraphics[width=\columnwidth]{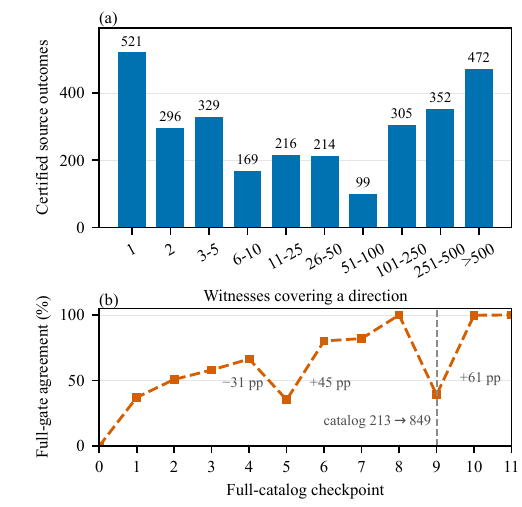}
\caption{(a) Covered source outcomes by number of covering witnesses. (b) Java agreement at each replay checkpoint, ending at 849/849. The dashed line marks catalog growth from 213 to 849 witnesses, not a regression.}\label{fig:progress}
\end{figure}

\subsection{Fault Detection (RQ3)}\label{sec:faults}
To verify our method's ability to detect unwanted silent business logic changes the agent could potentially introduce, we derived ten mutation operators based on Fig.~\ref{fig:defects}, limited to ten by the cost of finding reliable locations where these operators reside. Our coding agent then created a site search function that looks up the generated Java code to list where every operator can apply (a total of 27,218 candidate sites). We then used a fixed seed to pick 45 sites per operator, a total of 450 sites. Each version of the compiled Java code contained a single mutation. We then replayed the witnesses that reach each mutated line, while $\Phi_H$ stayed fixed. Mutants were killed if disagreement was found under $\Phi_H$. \texttt{REDEFINES} layout mutants have no executable line: they replay all 849 witnesses and count as reached if a witness enters a paragraph that names the overlaid field or its group, so this rule affects only the reached/unreached split. The gate killed 214 of 450 mutants (Table~\ref{tab:mutation}), 206 of them in code the witnesses reach. No rounding or precision mutant was killed. Of 32 sampled survivors, 6 were equivalent and 26 needed inputs that no witness reaching the site carries.

\begin{table}[t]
\caption{Seeded Java-only mutants, 45 per operator. Reached: a certified
witness executes the mutated line (for overlays, enters a paragraph naming the field). The gate also killed 8 unreached \texttt{REDEFINES} mutants (214 of 450 overall).}\label{tab:mutation}
\centering
\footnotesize
\begin{tabular}{@{}lrrr@{}}
\toprule
Operator & Reached & Killed & Rate \\
\midrule
Deleted \texttt{WRITE}/\texttt{REWRITE} & 43 & 43 & 100\% \\
Deleted \texttt{PERFORM} & 40 & 40 & 100\% \\
Negated condition & 42 & 39 & 93\% \\
Deleted assignment/\texttt{MOVE} & 38 & 26 & 68\% \\
Relational operator flip & 43 & 22 & 51\% \\
\texttt{REDEFINES} overlay offset & 19 & 6 & 32\% \\
Off-by-one \texttt{OCCURS} subscript & 37 & 11 & 30\% \\
AND/OR connector swap & 45 & 13 & 29\% \\
Packed-decimal sign & 37 & 6 & 16\% \\
Arithmetic precision/rounding & 35 & 0 & 0\% \\
\midrule
Total & 379 & 206 & 54\% \\
\bottomrule
\end{tabular}
\end{table}

\subsection{Repair Ledger (RQ4)}\label{sec:ledger}
\begin{figure}[t]
\centering
\includegraphics[width=\columnwidth]{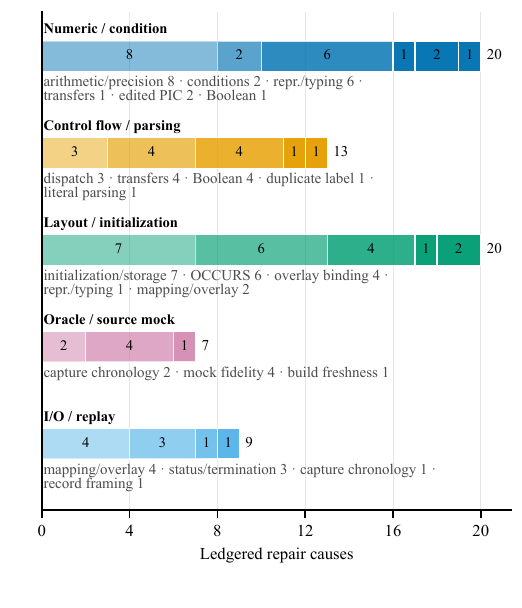}
\caption{Sixty-nine ledgered repair causes by semantic layer; annotations list each bar's mechanisms from left to right.}\label{fig:defects}
\end{figure}

We maintained a ledger to keep track of a filtered set of defects encountered
and repaired, in the Java or, when a disagreement traced back to them, in the surrogate, mock, or capture oracle, including deterministic disagreements that
were consistently recurring rather than one-offs, bugs that occurred
in shared code and thus affected multiple witnesses/branches, and
defects that were clearly identifiable rather than a collection of
small scattered tweaks. The final results can be seen in
Fig.~\ref{fig:defects}. Excluding transient bugs, the ledger holds 69 durable repairs: 61 in the Java and 8 in the surrogate, mock, or capture oracle, none of which touched the program's logic. The ledger was filtered and classified retrospectively, so its counts show which kinds of defects the method exposed, not a defect rate.

\subsection{Validation Loop and Cost}\label{sec:cost}
The Java replay used the chronological tape matcher in
Fig.~\ref{fig:tape-matcher} and the normalization contract in
Eq.~\eqref{eq:agreement}. As soon as $\Phi_H(E_J)$ diverged from
$\Phi_H(E_S)$ at any point, the agent would repair, recompile,
re-deploy the Java code and re-run the same witnesses until
$\Phi_H(E_J) = \Phi_H(E_S)$. Figure~\ref{fig:progress} shows the
dependency of progress on full catalog repairs, as previously passing
witnesses may fail post-repair. Table~\ref{tab:cost} reports the runtime of the final gate: about 12 minutes for all 849 witnesses.
Although we tracked the spend of 203 million tokens across 1,067 agent requests, these benchmarks cannot directly be attributed to algorithmic per-witness spend. The resources allocated included fixes to the reusable Java code generator that made it progressively better over time; it also included tape, runtime, COBOL mock, and parity repairs. Additionally, as the experiment spun across several weeks, the agent models were also uplifted, and hence the token consumption and caching methodologies, as well as costs, changed accordingly. 

\section{Discussion and Threats to Validity}\label{sec:discussion}

Certification guarantees repeatable source replay and target agreement under $\Phi_H$ only (RQ5). It does not guarantee production equivalence or any behavior outside $\Phi_H$.
As we only certified successful runs, error scenarios were excluded. Transient storage was also not tracked. To assess the risk of the coding agent introducing unintended business logic changes, we performed a mutation study (Section~\ref{sec:faults}) that showed 47.6\% fault detection. Our normalization process may also miss genuine parity failures. The GnuCOBOL compiler may not be 100\% identical to the original mainframe compiler; for example, ASCII versus EBCDIC mismatches or compiler defects may go unnoticed. For branching we used a limited set of probes: \texttt{IF}, \texttt{EVALUATE}, \texttt{PERFORM UNTIL}, \texttt{SEARCH}, and conditional handlers, which may not be a comprehensive list of inflection points. Future enhancements include a more comprehensive set of branching instrumentation, boundary-directed witness construction, and field-level payload comparison. Although our method does not guarantee 100\% code compatibility, we expect it to substantially reduce the number of defects prior to an integration testing phase.

As an engineering campaign that was carried out over time, it leveraged evolving agentic functionality that, in the course of the processing, fixed Java generator bugs, the mock layer, and the certification tools. As a result, it was impossible to establish a ``like-for-like'' equivalence with the Locksmith Loop witness method. Accordingly, we make no claim of a quantified improvement over the Locksmith Loop witness-generation approach; we report only the coverage achieved on this subject.

Entry-replay certification establishes that the COBOL surrogate and migrated Java agree for the certified witnesses under $\Phi_H$; it does not establish identical internal behavior.

\section{Conclusion}

We describe in this paper entry-replay certification, a novel method for gradually discovering repeatable entry witnesses starting with local isolated discovery and backward lifting. We were able to generate 849 certified witnesses covering 2,973/3,806, or 78.1\% of all source branches, and 989/1,033, or 95.7\% of the paragraphs, of a 33,429-line production-like COBOL program. A repair ledger supplied useful information about challenges encountered in this migration.
The gate exposed 61 Java defects during the campaign and detected 47.6\% of seeded mutants, but no rounding or precision defects, since the witnesses rarely exercise value boundaries.
We verified feasibility of the solution by creating repeatable source certification and reached COBOL-to-Java agreement for the certified witnesses under the declared observation projection.

\section*{Data Availability}
The COBOL source, fixtures, witnesses, and generated code are confidential.

\bibliographystyle{IEEEtran}
\bibliography{references}

\end{document}